\documentclass[12pt]{iopart}

\usepackage{iopams}  
\usepackage{graphicx}
\usepackage{braket}
\usepackage{siunitx}
\usepackage{color}
\usepackage[normalem]{ulem}

\begin{document}

\title[]{Phase response of atom interferometers based on sequential Bragg diffractions}

\author{B. D\'{e}camps, 
M. Bordoux,
J. Alibert,
B. Allard,
and A. Gauguet}

\address{Laboratoire Collision Agr\'{e}gats R\'{e}activit\'{e}, IRSAMC, Universit\'{e} de Toulouse, CNRS, UPS, Toulouse, France}
\ead{gauguet@irsamc.ups-tlse.fr}
\vspace{10pt}

\begin{indented}
\item[]
\date{\today}
\end{indented}

\begin{abstract}
Large Momentum Transfer (LMT) beam splitters are implemented in atom interferometers to increase their sensitivity. However, LMT-interferometer requires additional light-pulses that modify the response function of the atom interferometer. In this paper, we develop an analytical model for the sensitivity function of the LMT-interferometers using sequential accelerating light pulses. We use the sensitivity function to calculate the acceleration sensitivity taking into account the pulse duration. In addition, the sensitivity to laser phase fluctuations is calculated, and we show that the pulse sequence can be engineered to mitigate the phase noise sensitivity.
\end{abstract}

%
\vspace{2pc}

\submitto{\jpb}
\section{Introduction}
Light-pulse atom interferometers \cite{Borde89} are implemented for precision measurements in various areas \cite{FermiSchool}. In particular, they have been used for gravito-inertial measurements such as the Earth's gravitation \cite{louchet-chauvet_influence_2011,hu_demonstration_2013,freier_mobile_2016}, its gradients \cite{Sorrentino_13}, and rotations \cite{Gustavson_00,Berg_composite_2015,barrett_sagnac_2014}. They are also used for measuring fundamental constants such as the gravitational constant \cite{fixler_atom_2007, rosi_precision_2014}  or the fine structure constant \cite{bouchendira_new_2011, Parker_18}. In addition, atom interferometers with an increased sensitivity are potential candidates for laboratory tests of general relativity \cite{Dimopoulos_08}, weak equivalence principle \cite{Zhou_test_2013,Bonnin_13,Barrett_16,Overstreet_18}, abnormal acceleration at various length scale \cite{Wolf_07, Jaffe_17} or for gravitational waves detection \cite{canuel_matter-wave_2014}. In order to increase their sensitivity, a promising mean is to increase the momentum separation between the two arms of the interferometer.

Various solutions have been developed to implement Large Momentum Transfer (LMT) beam splitters in an atom interferometer: either by using a multi-photon transition with a single pulse \cite{muller_atom_2008}, or by using a beam splitter pulse ($\pi/2$-pulse) followed by an acceleration of the interferometer arms. The acceleration can be controled by an optical lattice (Bloch) \cite{McDo_faster_2014,clade_large_2009} or can result from a sequence of light pulses ($\pi$-pulse) \cite{chiow_102_2011}. These additional light pulses change the interferometer phase, which is why an accurate modeling of LMT atom interferometers response functions is required for precision measurements. Phase shift calculations for arbitrary atom interferometer geometries have been studied extensively \cite{Borde_02, Bongs_06,Dubetsky_06,Kleinert_15}. However, most of these methods are not convenient for modeling precisely the laser phase evolution experienced by the atoms during the light pulses.

In this paper, we calculate the response function of a LMT-interferometer based on the sequential light-pulse acceleration. The calculation relies on the sensitivity function formalism, initially developed for atomic clocks \cite{Dick_87}, which proved to be very efficient to analyse the sensitivity of light-pulse atom interferometers thereafter \cite{Cheinet08}. In particular this formalism can precisely calculate the impact of phase fluctuations at any frequency. It is also used to determine the modification of the interferometer space-time area due to the finite duration of the light pulses. A straightforward extension of this formalism can model any interferometers involving two quantum states. However, the LMT-interferometers based on sequential accelerating pulses couple more than two momentum states. In this paper, we extend the sensitivity function formalism to these LMT-interferometers and we derive an analytical solution for the response function in the temporal and the Fourier domain. The paper is structured as follows: In Sec.~\ref{sec2} we describe the atom interferometer modeled in this paper. In Sec.~\ref{sec3} we introduce the sensitivity function of this type of interferometer. In Sec.~\ref{sec4} we calculate the spectral response to evaluate phase noise contributions, emphasizing the specific features of the sequential LMT-interferometers. 
\section{The LMT atom interferometer}
\label{sec2}
\begin{figure}
\begin{center}
\includegraphics[width=0.49\textwidth]{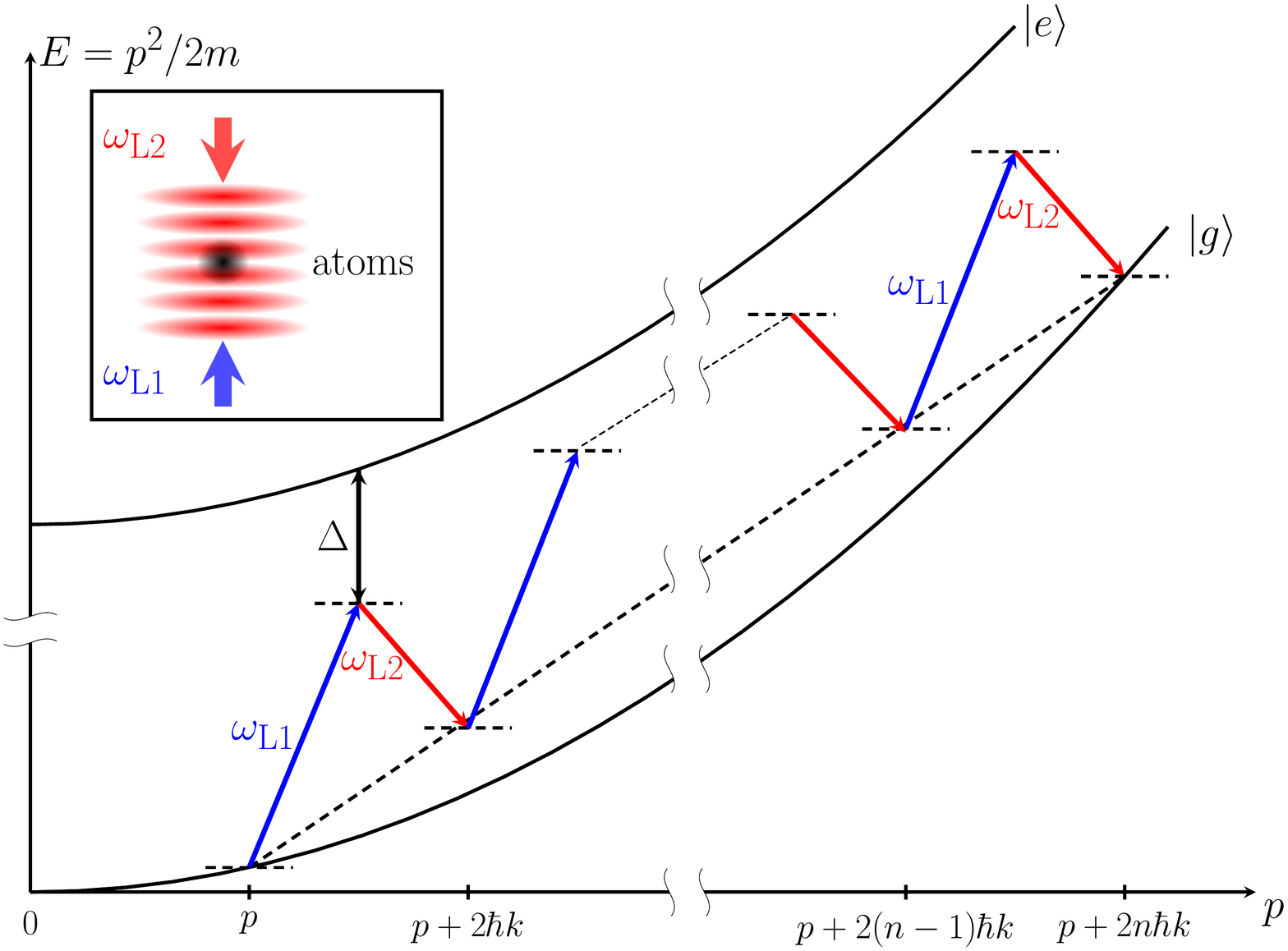}
\caption[bragg]{(Color online). High-order Bragg diffraction. The multiple interaction with the two far detuned Bragg beams at frequency $\omega_{L1}$ and $\omega_{L2}$ couple the momentum states $\ket{g,\mathbf{p}}$ and $\ket{g,\mathbf{p}+2 n \hbar \mathbf{k}}$. The diffraction order $n$ is set by the tuning of the laser frequencies. $\Delta$ is the single photon detuning with respect to the excited state $\ket e$. Inset : Counter-propagative Bragg configuration.\label{bragg}}
\end{center}
\end{figure}

We consider LMT atom interferometers based on sequential Bragg pulses. The Bragg lattices are created by two counter-propagative laser beams with adjustable frequencies $\omega_{\mathrm{L1}}$ and $\omega_{\mathrm{L2}}$ (see inset of figure~\ref{bragg}), and the phase difference between the two lasers is labelled $\phi$ in the following. The detuning $\Delta$ of the two beams with respect to the single-photon transition is large compared to the excited state line-width to avoid any spontaneous emission. 
 
Here, we consider high-order Bragg diffraction pulses that couple momentum states with a momentum separation of $2n \hbar \mathbf{k}$, where the order of diffraction $n$ is an integer number and $\mathbf{k}$ is the laser wave-vector. The high-order Bragg coupling, illustrated in figure~\ref{bragg} in the laboratory frame, can be modeled as an effective coupling between two momentum states $\ket{\mathbf{p}}$ and $\ket{\mathbf{p} + 2n \hbar \mathbf{k}}$. For a square pulse, the effective Rabi frequency between those two states $\Omega_{\mathrm{R}}$ is a function of the two-photon Rabi frequency $\Omega_0$ and the diffraction order $n$ \cite{Muller08}:
\begin{equation}
\label{Omegaeff}
\Omega_{\mathrm{R}} = \frac{\Omega_0^{n}}{[(n-1)!]^{2}}\frac{1}{\left(8 \omega_{\mathrm{r}}\right)^{n-1}},
\end{equation}
where $\omega_{\mathrm{r}} = \frac{\hbar \mathbf{k}^{2} }{2m}$ is the recoil frequency for an atom of mass $m$. The relevance of this equation is questionable for high $n$. However, it illustrates that the required laser intensity increases rapidly with the Bragg order $n$. In addition, for a $n-th$ order Bragg transition, the laser phase imprinted on the diffracted state $\ket{\mathbf{p} + 2n \hbar \mathbf{k}}$ is $n \times \phi$.

\begin{figure}[!htb]
\begin{center}
\includegraphics[width=0.49\textwidth]{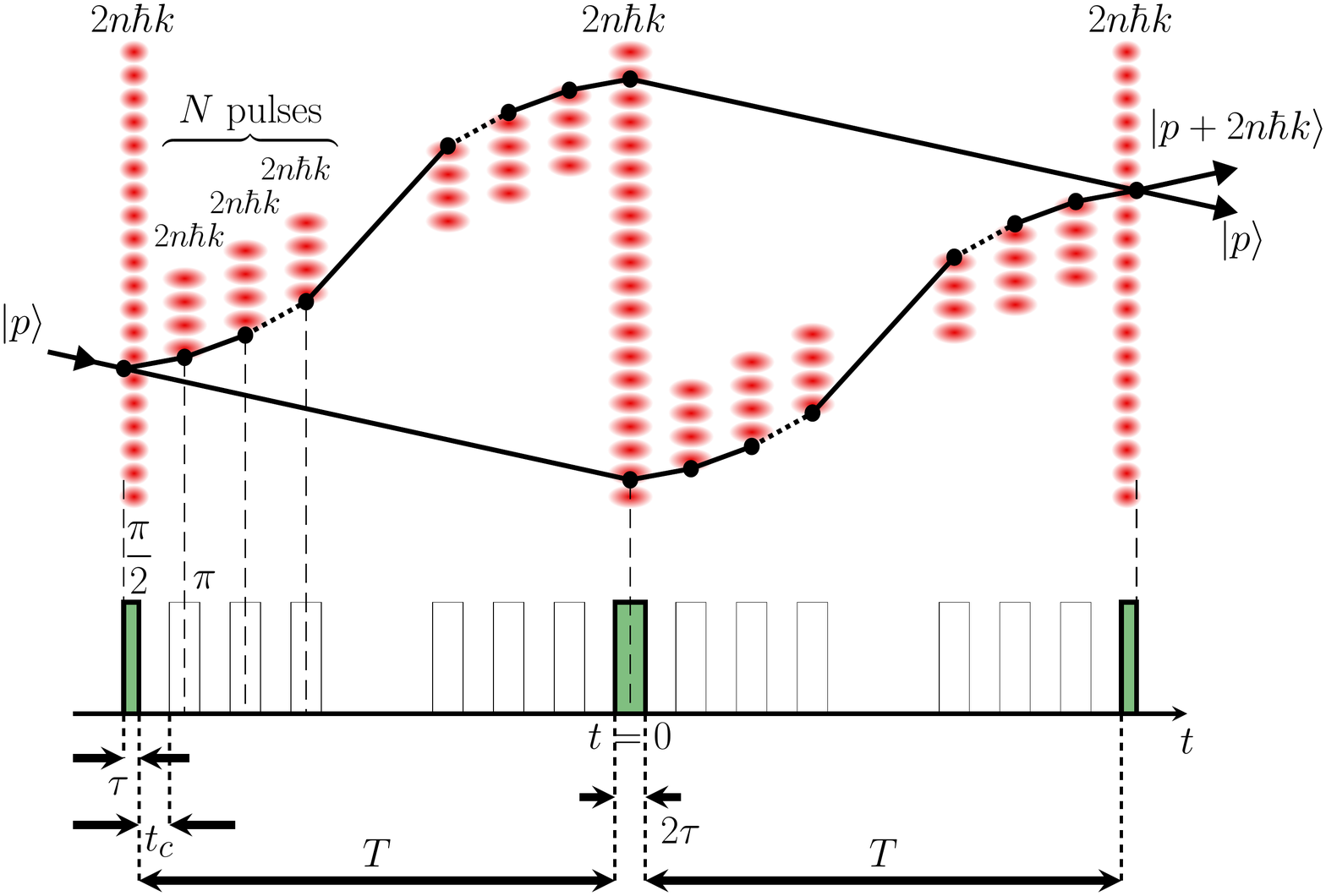}
\caption[interfero]{(Color online). The upper panel represents the space-time diagram of the considered LMT interferometer in the Bragg lattice frame. The three long red lattices are the diffracting $\pi/2$-$\pi$-$\pi/2$-pulses of a standard Mach-Zehnder atom interferometer with a $2T$ interrogation time. The space-time area is increased by sequences of $N$ accelerating $\pi$-pulses separated by a time $t_{c}$, acting only on a single arm (short red lattices). Each $n$-th  order Bragg pulse lasts $\tau$ (resp. $2\tau$) for $\pi/2$-pulses (resp. $\pi$-pulses). The lower panel sketches the intensity of the Bragg beams as a function of time.\label{interfero}}
\end{center}
\end{figure}

The atom interferometer considered is based on a Mach-Zehnder geometry as shown in figure~\ref{interfero}. It consists in a series of three diffracting Bragg pulses that act on both arms of the interferometer. The input state of the interferometer is $\ket{\mathbf{p}}$. A first $\pi/2$-pulse of duration $\tau$ creates a coherent superposition of the two momentum states $\ket{\mathbf{p}}$ and $\ket{\mathbf{p}+ 2 n\hbar \mathbf{k}}$. After a time $T$, a central $\pi$-pulse of duration $2\tau$ exchanges the momentum of the two arms. Finally, after another time $T$, the last $\pi/2$-pulse closes the interferometer paths. Those diffracting pulses are illustrated in figure~\ref{interfero} by long red lattices. 

To increase the momentum separation between the two interferometer arms, accelerating pulses are resonant with only one of the interferometer arms at a time. After the first diffracting pulse, a first sequence of $N$ $\pi$-pulses of duration $2\tau$ transfers $N\times 2n\hbar \mathbf{k}$ to the upper arm. The accelerating pulses are regularly separated by a time $t_{c}$. For simplicity, we consider that the delay between the end of the first diffracting $\pi/2$-pulse and the first accelerating pulse is $t_{c}$. A second sequence of $N$ $\pi$-pulses decelerates the upper arm down to $\ket{\mathbf{p} + 2 n\hbar k}$ before the central diffracting $\pi$-pulse. It is important to note that each accelerating pulse of the first sequence has its symmetric decelerating pulse in the second sequence. The same acceleration-deceleration sequence is applied to the lower arm after the central $\pi$-pulse. 

The interferometric signal corresponds to the probability of finding the atoms in either state at the output port of the interferometer. It is determined from the atomic populations measured in each interferometer outputs:
\begin{equation}
P = \frac{N_{\ket{\mathbf{p}}}}{N_{\ket{\mathbf{p}+ 2 n\hbar \mathbf{k}}}+N_{\ket{\mathbf{p}}}}= \frac{1+\cos\Phi}{2},
\end{equation}
with $\Phi$ the phase difference between the two arms that contains the signal of interest. In many cases, the measured phase shift $\Phi$ is proportional to the effective momentum separation $\mathbf{k}_{\mathrm{eff}}= 2 n (N+1) \mathbf{k}$. 
 
With uncorrelated atoms, the smallest measurable phase shift (the phase sensitivity) is ultimately limited by the quantum projection noise that scales with the square root of the detected atom number \cite{Itano93}. In practice, the phase sensitivity is lowered by laser phase noises that can be evaluated with the sensitivity function.
\section{Sensitivity function}
\label{sec3}
An infinitesimal laser phase shift between the two lasers of the Bragg lattice $\delta \phi$ modifies the atom interferometer phase shift $\Phi$ and so the population measured at the interferometer outputs. The sensitivity function corresponds to the phase response of the interferometer for a laser phase shift $\delta \phi$ occurring at time $t$. The sensitivity function $g_{\phi}(t)$ of the interferometer is defined by:
\begin{equation}
\label{def-gphi}
g_{\phi}\left(t\right) := \lim \limits_{\delta \phi \rightarrow 0} \frac{\delta \Phi\left(\delta \phi,t\right)}{\delta \phi}=\frac{2}{\sin\Phi} \lim \limits_{\delta \phi \rightarrow 0} \frac{\delta P\left(\Phi,\delta \phi,t\right)}{\delta \phi}
\end{equation}
This function is used to determine the interferometric phase shift $\Phi$ for a laser phase fluctuation $\phi(t)$ during the interferometer sequence:
\begin{equation}
\label{Phi-gphi}
\Phi = \int_{-\infty}^{+\infty} g_{\phi} \left(t'\right)\frac{d\phi}{dt}\left(t'\right)dt'
\end{equation}
Our calculation of the sensitivity function relies on three main assumptions. First, we model each multi-photon transitions with an effective coupling between only two states. Second, the square light pulses considered are on resonance with the effective transition. Third, the interferometer is operated at its maximal sensitivity $\Phi \sim \pi/2$.
\subsection{Sensitivity function for a 3-pulses interferometer.}
\begin{figure}
\begin{center}
\includegraphics[width=0.5\textwidth]{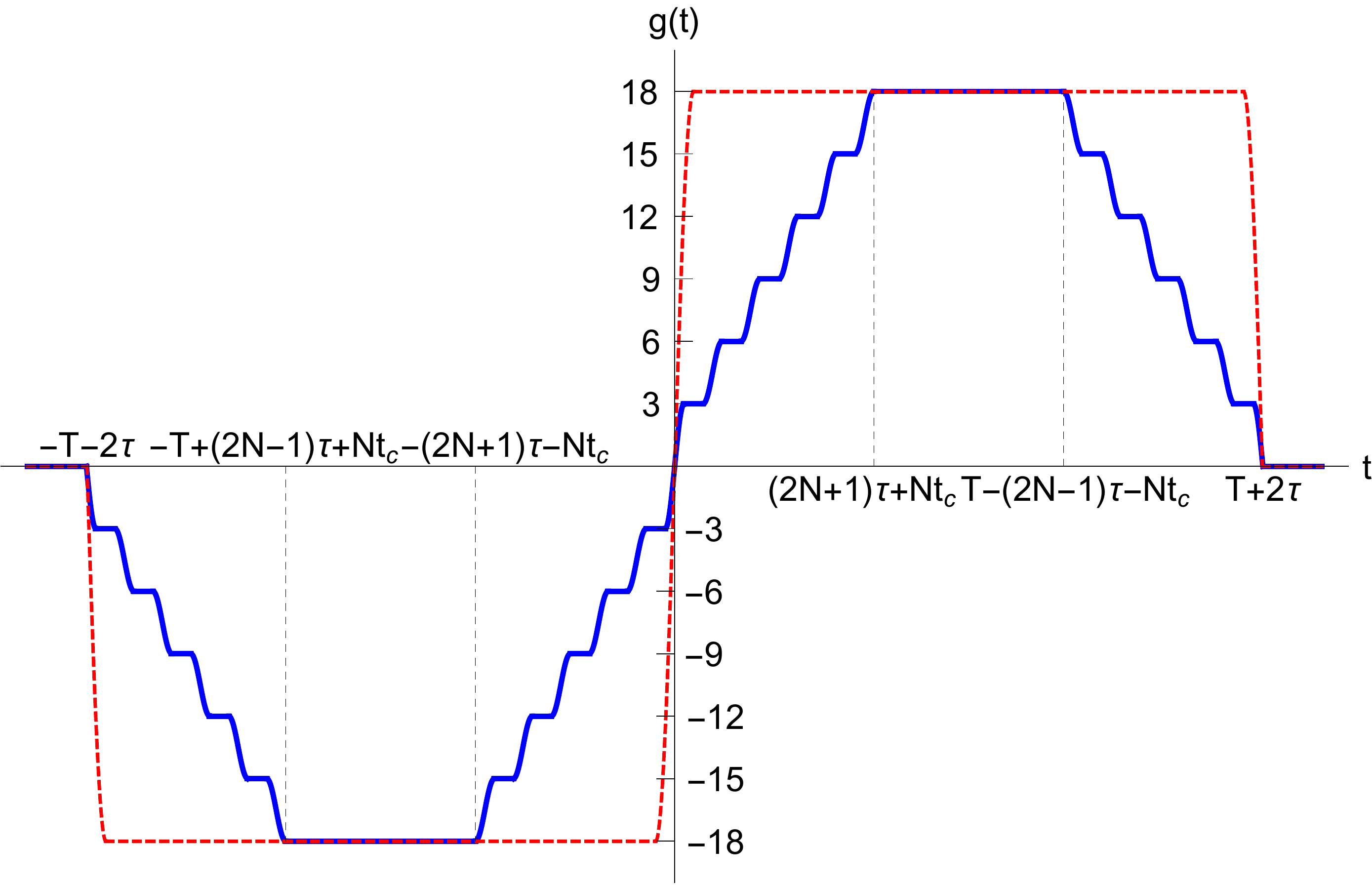}
\caption{(Color online). Sensitivity functions of a 3-pulses interferometer (dashed red) and N=5 LMT-interferometer (continuous blue). 
In order to compare similar scaling factor, the diffraction order is $n=18$ for the 3-pulses interferometer and $n=3$ for the LMT-interferometer sequence. \label{fig.Sensifunc}}
\end{center}
\end{figure}

The sensitivity function of the standard 3-pulses interferometer has been calculated in \cite{Cheinet08} for the effective 2-levels system associated with a two-photon Raman transition \cite{Kasevich92}. Their results can be directly adapted to any effective 2-level systems. In particular, we consider the coupling between two momentum states for an atom in an optical lattice in the Bragg regime. The sensitivity function $g^{\left(0\right)}_{\phi}\left(t\right)$ is an odd function of time so it is completely determined for $t \geqslant 0$:
\begin{equation}
\label{gphi-3pulses}
g^{\left(0\right)}_{\phi}\left(t\right) = 
\left \{
\begin{array}{c l}
n \times \sin\left(\Omega_{\rm{R}} t\right) & t \in \left[0;\tau\right[\\
n & t \in \left[\tau;T+\tau\right[\\
- n \times \sin\left(\Omega_{\rm{R}} \left(T-t\right)\right) & t \in \left[T+\tau;T+2\tau\right]
\end{array}
\right.
\end{equation}
The sensitivity function is plotted in figure~\ref{fig.Sensifunc} for an interferometer with momentum separation of $36 \hbar k$ using high order Bragg diffraction $\left(n=18\right)$.
\subsection{Sensitivity function for a sequential LMT-interferometer.}
\begin{figure}
\begin{center}
\includegraphics[width=0.5\textwidth]{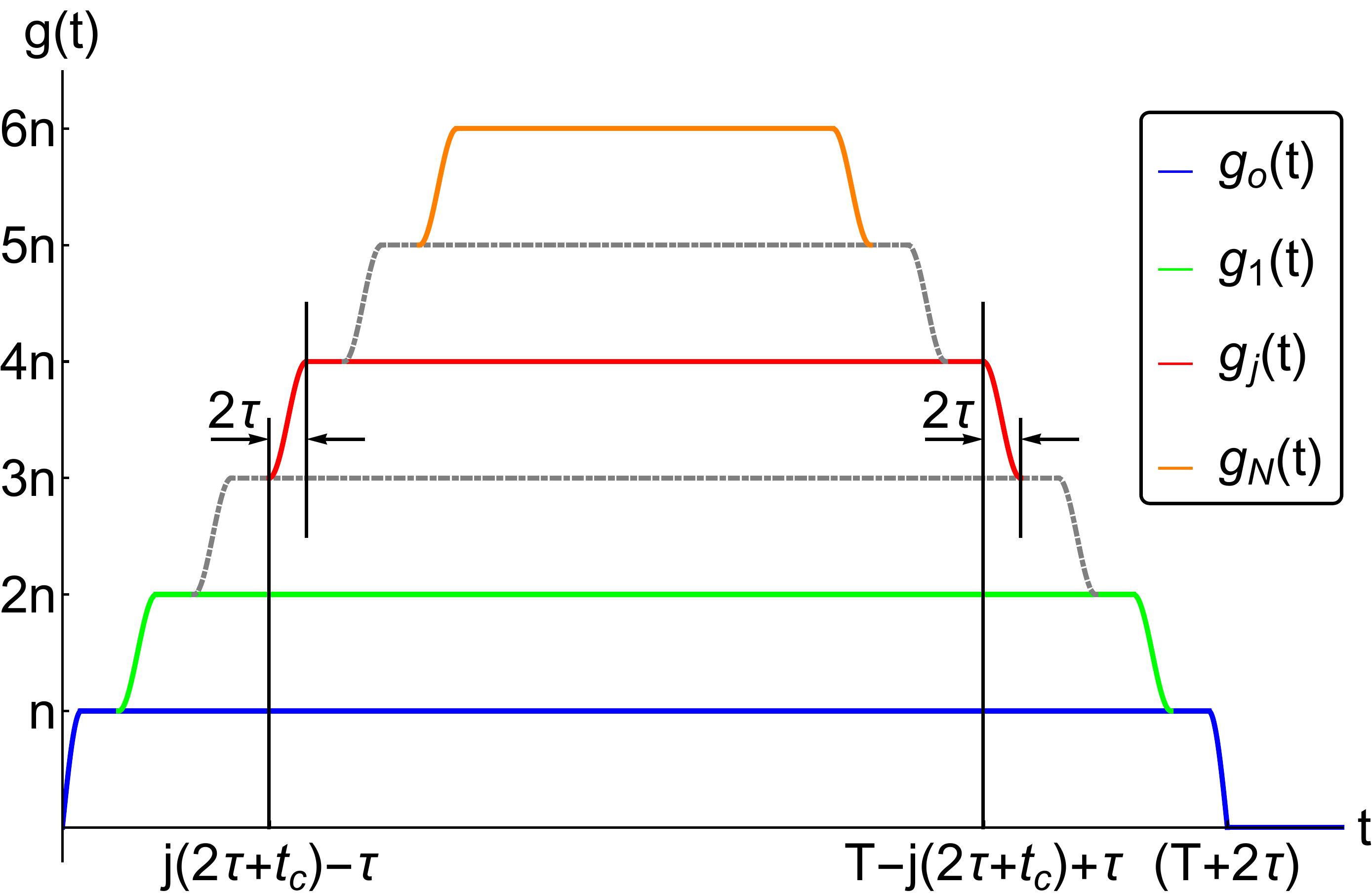}
\caption{(Color online). Representation of the partial sensitivity function as defined in equation (\ref{gk}) for $N=5$. The timing of the j-th pulse (countinuous red line) as well as the end of the final pulse are indicated on the temporal axis.\label{fig.partialSensi}}.
\end{center}
\end{figure}

This paper focuses on the extension of the previous formalism in order to include the phase shift induced by the additional accelerating light pulses. The calculation of the sensitivity function for a sequential LMT-interferometer $g^{\left(LMT\right)}_{\phi}$ is detailed in the \ref{app1}. The derivation is based on the separation of $g^{\left(LMT\right)}_{\phi}$ into a discrete sum of terms with different starting times and durations (see figure~\ref{fig.partialSensi}). 

The first term $g_{0}\left(t\right)$ corresponds to the sensitivity function of the diffracting pulses $\left(\pi/2-\pi-\pi/2\right)$ interacting with the two arms of the interferometer. With our choice of timings (total duration from the first pulse to the last pulse equal to $2\left(T+2\tau\right)$) it is equal to the 3-pulses sensitivity function: 
\begin{equation}
\label{g0}
g_{0}\left(t\right) = g^{\left(0\right)}_{\phi}\left(t\right)
\end{equation}

We define a sensitivity function $g_{j}\left(t\right)$ for each pair of accelerating and decelerating pulses between the momentum states $j\times 2n\hbar \mathbf{k}$ and $\left(j+1\right)\times 2 n \hbar \mathbf{k}$ and we find that:
\begin{equation}
\label{gk}
g_{j}\left(t\right) = \left \{
\begin{array}{c l}
0 & t \in  \left[0; t_{j1}\right[\\
n \times \sin\left(\frac{\Omega_{\rm{eff}} \left(t-t_{j1}\right)}{2}\right)^{2} & t \in  \left[t_{j1};t_{j2}\right[ \\
n & t \in  \left[t_{j2};t_{j3}\right[\\
n \times \sin\left(\frac{\Omega_{\rm{eff}} \left(t_{j4}-t\right)}{2}\right)^{2} & t \in  \left[t_{j3};t_{j4}\right[\\
0 & t \in  \left[t_{j4};T+2\tau\right]
\end{array}
\right.
\end{equation}
The $j-th$ accelerating pulse after the diffracting $\pi$-pulse starts at $t_{j1} = j\left(2\tau+t_{c}\right)-\tau$ and ends at $t_{j2}= j\left(2\tau+t_{c}\right)+\tau$, it is combined with the $j-th$ decelerating pulse which starts at $t_{j3}= T-j\left(2\tau+t_{c}\right)+\tau$ and ends at $t_{j4}= T-j\left(2\tau+t_{c}\right)+3\tau$. The global sensitivity function is then:
\begin{equation}
\label{gLMTphi}
g^{\left(LMT\right)}_{\phi}\left(t\right) = \sum_{k=0}^{N} g_{j}\left(t\right)
\end{equation}
The sensitivity function for a LMT-interferometer with momentum separation of $36 \hbar k$ is shown in the figure~\ref{fig.Sensifunc} in case of $N=5$ and $n=3$.
\subsection{The acceleration scale factor}
The phase shift of a LMT-interferometer can be calculated from a deterministic laser phase evolution $\phi\left(t\right)$ experienced by the atom by integrating the sensitivity function ( \ref{Phi-gphi}). We consider atomic motion described by a constant acceleration $\mathbf{a_c}$ in the laboratory frame. Therefore an atom moving in the laser lattice experiences a laser phase evolution given by:
\begin{equation}
\label{eq.acc}
\frac{d \phi \left(t\right)}{dt} = 2 \;\mathbf{k \cdot a_{c}} \; t + 2 \frac{\mathbf{k \cdot p}}{m}
\end{equation}
In practice the frequency difference between the lasers is adjusted to keep the Bragg resonance condition during all the light-pulses, according to the assumption made for the calculation. One can compute directly the phase resulting from the constant acceleration which accounts for the finite duration of the light-pulses. As the sensitivity function is anti-symmetric, the integral over $g_{\phi}\left(t\right) p$ vanishes, which means that this atom interferometer geometry is independent of the initial atom momentum $\mathbf{p}$. The phase shift obtained for a 3-pulses sequence $\left(N=0\right)$ is:
\begin{equation}
\label{eq.phiacc3p}
\begin{array}{cll}
\frac{\Phi_{\rm{acc,3}}}{\mathbf{k_{\mathrm{eff}}} \cdot \mathbf{a_{c}}T^2} & = & 1+\left(1+\frac{2}{\pi}\right)\frac{2\tau}{T} + \frac{2}{\pi} \left(\frac{2\tau}{T}\right)^{2}
\end{array}
\end{equation}
The phase shift obtained for LMT-interferometer is:
\begin{equation}
\label{eq.phiacc}
\begin{array}{cll}
\frac{\Phi_{\rm{acc}}}{\mathbf{k_{\mathrm{eff}}} \cdot \mathbf{a_{c}}T^2} & = & 1+\left(1+\frac{2}{\pi}-N\left(N-1\right)\right)\frac{1}{N+1}\frac{2\tau}{T} - N \frac{t_c}{T} \\[0.3cm]
&& + \left(\frac{2}{\pi}+N\right)\frac{1}{N+1}\left(\frac{2\tau}{T}\right)^2

\end{array}
\end{equation}

In order to compare the sensitivity between the N-pulses and the standard 3-pulses interferometers, we consider identical pulse durations $\left(\tau\right)$ and an identical total momentum transfer $\mathbf{k}_{\mathrm{eff}}$. In the limit of infinitely short light-pulses ($t_{c}$, $\tau \rightarrow 0$), we find the expected factor which scales as the momentum difference and the interferometric time squared $k_{\mathrm{eff}} T^{2}$. However, precision measurements need an accurate knowledge of the scale factor which requires to consider the additional terms in $\tau/T$ and $t_{c}/T$. In the case of the LMT sequence, momentum transfer takes a longer duration than the equivalent 3-pulse interferometer and this slightly modifies the space-time area:
 \begin{equation}
\label{eq.phiacc-phiacc3}
\begin{array}{cll}
\frac{ \Phi_{\rm{acc,3}}-\Phi_{\rm{acc}}}{\mathbf{k_{\mathrm{eff}} \cdot a_{c}}T^2} &=& \frac{N}{N+1} \left(\frac{2}{\pi}+N\right) \frac{2 \tau}{T} +N \frac{t_c}{T}\\[0.3cm]
&&-\frac{N}{N+1} \left( 1 - \frac{2}{\pi} \right) \left(\frac{2 \tau}{T} \right)^2
\end{array}
\end{equation}
\section{Phase noise sensitivity} 
\label{sec4}
To estimate the impact of the laser phase noise on the interferometer sensitivity, it is convenient to define the transfer function of the interferometer in the Fourier domain $H\left(\omega\right)$. The variance of the interferometric phase is given by the power spectral density $S_{\phi}\left(\omega\right)$ of the laser phase noise weighted by the transfer function:
\begin{equation}\label{eq.laserphasenoise}
\sigma_{\Phi}^2  = \int_{0}^{\infty} \left \vert H\left(\omega\right)\right \vert^{2} S_{\phi}\left( \omega \right) \frac{d \omega}{2 \pi}
\end{equation}
With the transfer function of the interferometer $H\left(\omega\right)$ defined as:
\begin{equation}
H\left(\omega\right) := \omega \int_{-\infty}^{+\infty}e^{-i \omega t}g_{\phi}\left(t\right) dt
\end{equation}
To determine $H\left(\omega\right)$, we define the partial transfer function for each individual partial sensitivity function $g_{j}\left(t\right)$:
\begin{equation}
\label{Hkdef}
H_{j}\left(\omega\right) := \omega \int_{-\infty}^{+\infty}e^{-i \omega t}g_{j}\left(t\right) dt
\end{equation}
From the equations (\ref{gk}) and (\ref{Hkdef}) we calculate each partial transfer function. For $j=0$, we recover the usual 3 pulses transfer function \cite{Cheinet08}:
\begin{equation}
\label{H0}
\begin{array}{cll}
H_{0}\left(\omega\right) &=& \frac{4 i n \Omega_{R}^{2}}{\omega^{2}-\Omega_{R}^{2}} 
\sin\big(\omega \frac{T+2\tau}{2}\big)\times\\[0.2cm]
 && \left\{ \sin\big(\omega \frac{T}{2}\big)+\frac{\omega}{\Omega_{R}} \cos \big(\omega \frac{T+2\tau}{2}\big)\right\}
 \\[0.3cm]
\end{array}
\end{equation}
\noindent For $j \neq 0$, the partial transfer functions is given by:
\begin{equation}
\label{Hk}
\begin{array}{cll}
H_{j} (\omega) &=& \frac{4 i n \Omega_{R}^{2}}{\omega^{2}-\Omega_{R}^{2}} 
\sin\big(\omega \frac{T+2\tau}{2}\big)\times\\[0.2cm]
 && \sin\big(\omega \frac{T-2j\left(2\tau+t_{c}\right)+2\tau}{2}\big) \cos \left(\omega \tau\right)
 \\[0.3cm]
\end{array}
\end{equation}

The transfer function of the interferometer $H\left(\omega\right)$ is the sum of all the partial transfer function. First we calculate the sum $\sum_{j=1}^{N} H_{j}\left(\omega\right)$, dropping the global factor independent on $j$, one gets:
\begin{equation}
\label{sumk}
\sum_{j=1}^{N} H_{j}\propto\sum_{j=0}^{N-1} \sin \bigg( \omega \frac{T-2\left(\tau+t_{c}\right)+j\left(-4 \tau-2t_{c}\right)}{2} \bigg)
\end{equation}

\noindent This expression is simplified with the following property \cite{Alan08}:
\begin{equation}
\sum_{j=0}^{N-1}\sin\left(x+jy\right)=\frac{\sin\left(x+\frac{\left(N-1\right)y}{2}\right)\sin(\frac{Ny}{2})}{\sin\left(y/2\right)}
\end{equation}

\noindent Finally, we get the global transfer function of the atom interferometer, which is the central result of this paper:
\begin{eqnarray}
\label{H}
H\left(\omega\right)=&& \frac{4 i n \Omega_{R}^{2}}{\omega^{2}-\Omega_{R}^{2}} \sin\left(\omega \frac{T+2\tau}{2}\right) \!\! \times \!\!  \nonumber \\[0.3cm]
&&\left\{\!\frac{\omega}{\Omega_{R}}\! \cos\left(\omega \frac{T+2\tau}{2} \right)+\sin\left(\omega \frac{T}{2}\right)+\right. \\[0.3cm]
&&\hspace{-1.3cm} \frac{\sin\left(\omega \frac{N(t_{c}+2\tau)}{2}\right)}{\sin\left(\omega \frac{t_{c}+2\tau}{2}\right)}\cos\left(\omega \tau\right)\sin\left(\omega \frac{T-(N+1)t_{c}-2N\tau}{2}\right)
\bigg\}  \nonumber \\[0.3cm]
\nonumber 
\end{eqnarray}

\begin{figure}
\begin{center}
\includegraphics[width=0.5\textwidth]{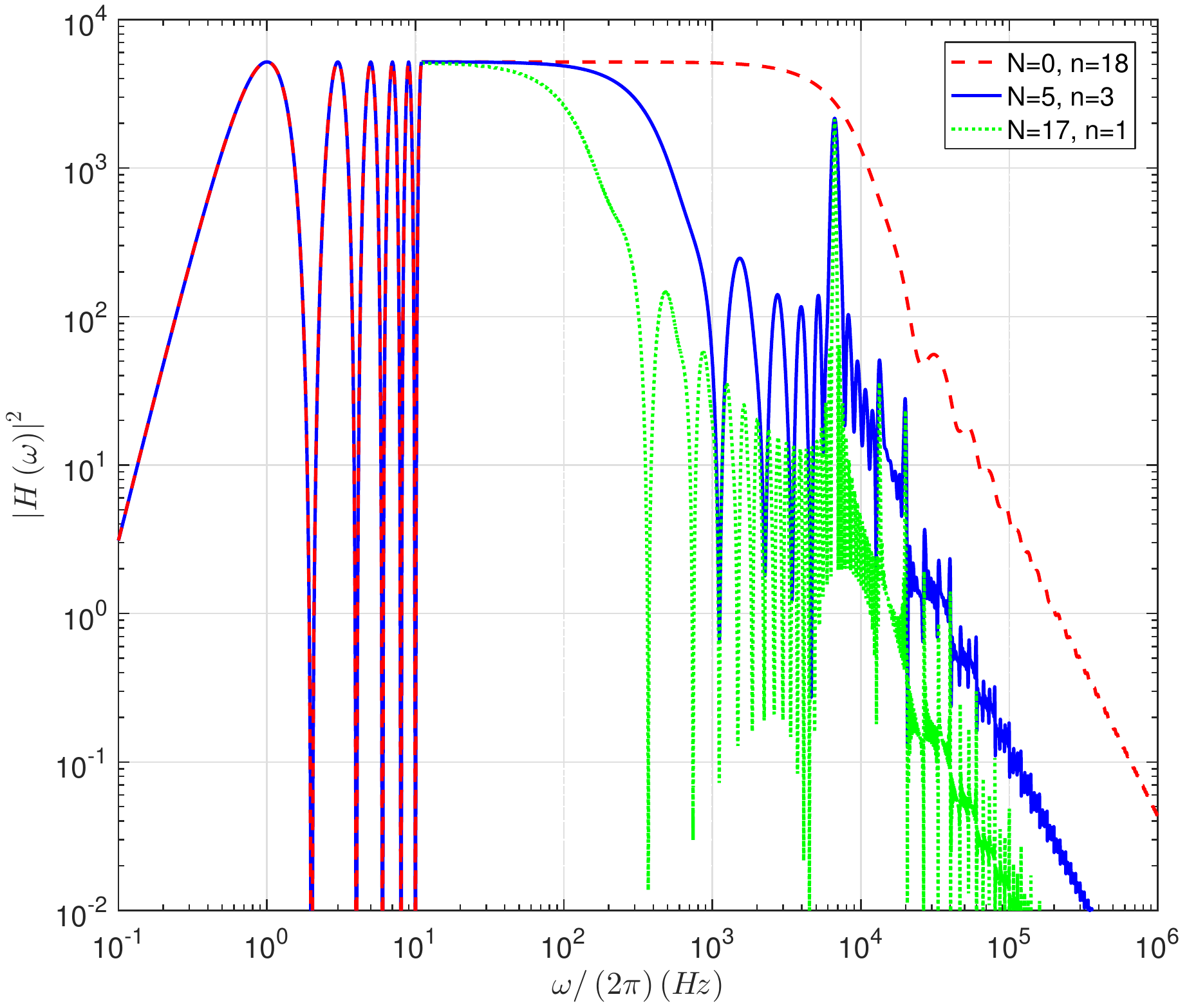}
\caption{(Color online). Transfer functions $| H\left(\omega\right) |^2$ of a 3-pulse (red dashed line), N=5 multi-pulses (blue solid line) and N=17 multi-pulse (green dashed-dotted line), for $2T=\SI{500}{\milli \second}$, $\tau=\SI{50}{\micro \second}$ and $\tau_{c}=\SI{50}{\micro \second}$.}
\label{fig.Transfunc}
\end{center}
\end{figure}
In figure~\ref{fig.Transfunc}, we plot $|H\left(\omega\right)|^2$ for three interferometers of identical space-time area ($2T=\SI{500}{\milli \second}$, and 36 $\hbar k$). We consider a 3-pulse interferometer with high-order Bragg pulses $n=18$ (red dashed line), a LMT-interferometer with first-order Bragg pulses $n=1$ and many accelerating pulses $N=17$ (green dashed-dotted line) and an intermediate configuration with $N = 5$ and $n=3$ (blue solid line). In addition, we keep the pulse duration $\tau = \SI{50}{\micro \second}$, and the pulse delay $t_{c} = \SI{50}{\micro \second}$ indentical for the three configurations. 

At low frequency ($\omega/(2 \pi)< (2N+1)/(2T)$), $|H\left(\omega\right)|^2$ does not depend on the specific sequence of the beam splitters. In particular, the finite duration $2T$ of the interferometer induces periodic cancellation at frequencies of $1/\left(T+2\tau\right)$ (2 Hz with our parameters). This feature is shown in figure~\ref{fig.Transfunc} at low frequencies ($f < 11$ Hz). At higher frequencies ($>11$ Hz), we plot the averaged value of $|H\left(\omega\right)|^2$ over a $1/T$ frequency span for a better readability of its envelope. Besides, a numerical factor multiply the averaged plots to match the low frequency maxima obtained without averaging.

The finite duration of the pulses leads to a low-pass filtering with a $1/f^2$ scaling which is characteristic of rectangular pulses \cite{Fang_18}. In the case of the 3-pulse interferometer the cut-off frequency is $\sim 1/\left(2 \tau \right)$ (10 kHz with our parameters). For the LMT-interferometer, the effective duration of the beam splitter leads to a lower cut-off frequency $\sim (2N\times(2 \tau + t_c))^{-1}$. Moreover, it is interesting to note that, regardless of the partial cancellation of sensitivity to frequencies above the effective cutoff, a residual phase sensitivity around $\omega/(2 \pi) \sim 1/(2 \tau+ t_c)$ remains independently of the number of intermediate pulses.

\begin{figure}
\begin{center}
\includegraphics[width=0.48\textwidth]{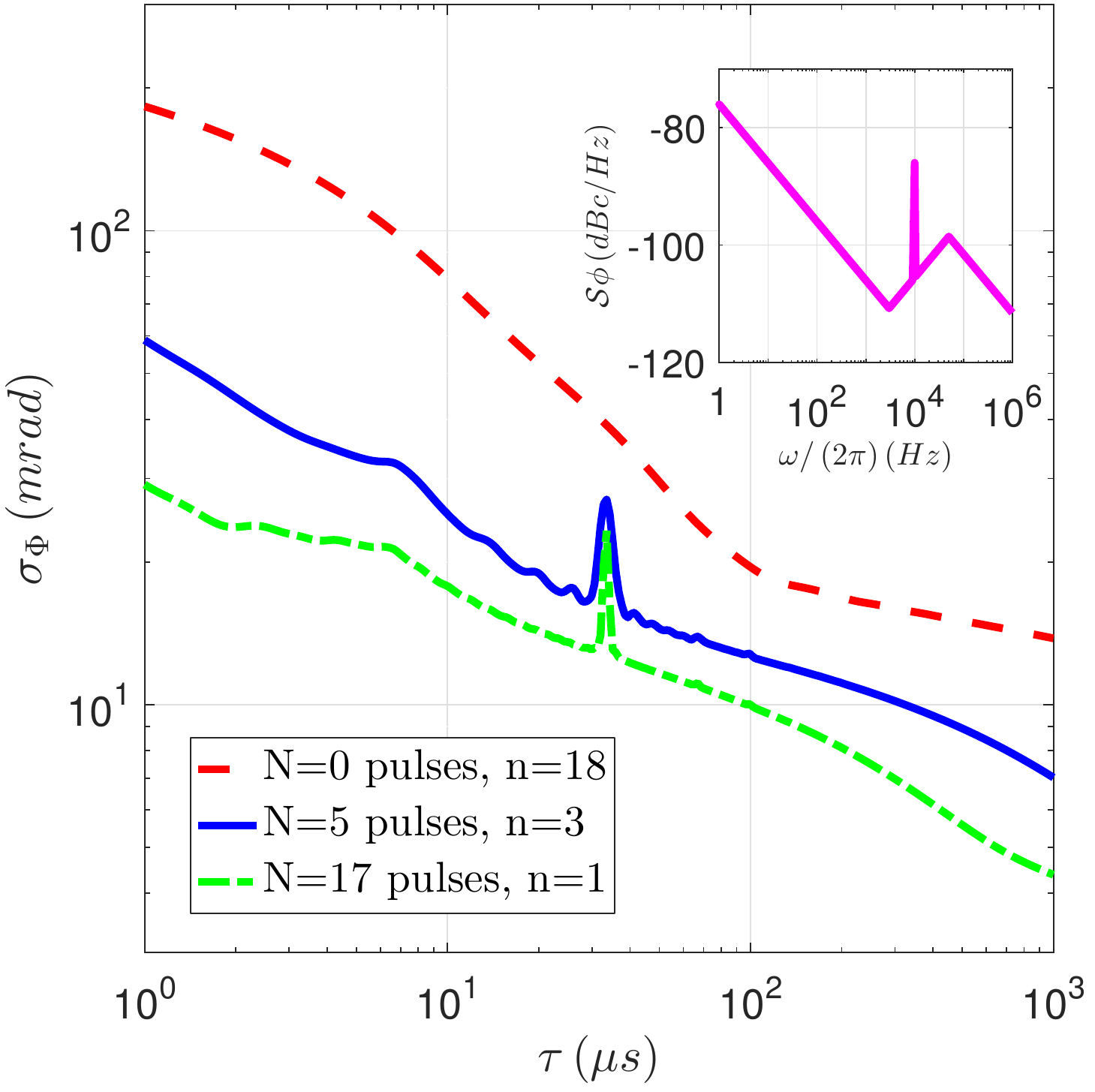}
\caption{(Color online). Interferometer phase noise from equation (\ref{eq.laserphasenoise}) for different diffracting pulses durations. The interferometric time is constant ($2T=\SI{500}{\milli \second}$) and for simplicity, the delay between each pulses is chosen equal to the pulse duration ($t_{c}=\tau$). The phase power spectrum density used is shown as an inset. Different interferometer configurations are considered (N=0, red dashed line, N=5, blue solid line and N=17, green dot-dashed line.)} 
\label{fig.Phase_noise}
\end{center}
\end{figure}

In order to illustrate the impact of the pulse sequence, we modeled a laser phase noise characterized by the power spectral density $S_{\phi}\left(\omega\right)$ shown in the inset of the figure~\ref{fig.Phase_noise}. In this example, we considered phase-locked lasers shifted by a few MHz for Bragg diffraction. The phase noise is representative of optical phase locked loop limited by the finite bandwidth of acousto-optic modulators used for the feedback ($\sim 50$kHz in our example). The figure \ref{fig.Phase_noise} shows the laser phase noise according to the pulse duration $\tau$ for the three interferometer configurations (N=0, N=5 and N=17). It emphasizes the decrease in phase noise reachable in a multi-pulse configuration due to the high frequency filtering. Moreover, we added a sharp resonance in the noise spectra  ($f_0 = 10$~kHz in our example), which leads to an increased phase noise for a pulse duration of $\sim 1/(3 f_0)$.  This example shows the possibility of the developed formalism to design an optimized interferometer sequence depending on the specific technical laser phase noise and/or power limitation.

\section{Conclusion}
We derived the response function of a light-pulse atom interferometer using a sequential diffraction in the temporal and in the Fourier domain.  The model presented in this paper is based on sequential effective two-level couplings, hence it can be extended to the other sequential light pulses such as two-photon Raman transition or single-photon transition.The precise knowledge of this transfer function allows to calculate the influence of several noise contributions onto the atom interferometers. In particular, we highlighted resonances in the transfer function depending on the sequence of the LMT beam splitters. In addition, a central result of our study is the rejection of high frequency laser phase fluctuations for sequential pulses LMT beam splitters. The modification of the response function is also of great interest for implementing an active noise rejection with an external vibration sensor. We also presented the modification of the space-time area of the interferometer that impacts on the scale factor of inertial sensor. Future work should implement the combination of pulse shaping \cite{Fang_18} and the multi-pulses configuration studied in this paper, which would be directly applicable to future experiments using sequential LMT-interferometers. Finally, the sensitivity function can be used for the analysis of systematic effects induced by the light pulses including light-shift which are of importance for LMT-interferometers. In this prospect, it would be of interest to consider the case of non-resonant accelerating pulses and the effect of the multi-photon light shifts \cite{Gauguet_08} during the acceleration pulses.
\ack
This work has been partially supported through the grant NEXT ANR-10-LABEX-0037 in the framework of the ``Programme des Investissements d'Avenir'', the French space national agency CNES, the Universit\'{e} F\'{e}d\'{e}rale de Toulouse Midi-Pyr\'{e}n\'{e}es through the program emergence-CORSAIR and the program Equipement-Ultitech-IRSAMC. J. A. and M. B. acknowledge support from CNES and the R\'{e}gion Occitanie.
\appendix
\section{Detailed derivation of the sensitivity function}
To compute the LMT-interferometer sensitivity function we use a formalism based on a matrix formulation of the momentum state superposition during the interferometer sequence. For a N-pulse sequence, we consider momentum states $ \left \lbrace \left \vert j \right \rangle := \left \vert \mathbf{p}+ j \times 2n\hbar \mathbf{k} \right \rangle\right \rbrace _{j \in \left[0,N\right]}$ where $\mathbf{p}$ is the initial atomic momentum. From an initial wave function $\left \vert \Psi_{0}\right \rangle = \left \vert 0\right \rangle $ one can compute the atomic wave function evolution by considering a sequence of light pulses  coupling only neighboring states and leaving the other states evolving freely. Each step can be represented by a matrix $M_{j}$ which is block diagonal with a unique $2 \times 2$ irreducible subspace coupling $\left \vert j \right \rangle$ and $\left \vert j+1\right \rangle$ expressed as:

\begin{equation}
\!\!\!\!\!\!\!\!\!\!\!\!\!\!\!\!\!\!\!\!\!\!\!\!\!\!\!\!\!\!\!\!\!\!\!\!
\small
\left(
\begin{array}{cc}
\cos\left[\frac{\Omega_{\mathrm{R}} \left(t_{f}-t\right)}{2}\right] e^{-i \omega_{j} \left(t_{f}-t\right)} & -i e^{i \left(\omega_{12} t + n \phi\right)} \sin \left[\frac{\Omega_{\mathrm{R}} \left(t_{f}-t\right)}{2}\right] e^{-i \omega_{j} \left(t_{f}-t\right)} \\
- i e^{-i \left(\omega_{12} t + n \phi\right)} \sin\left[\frac{\Omega_{\mathrm{R}} \left(t_{f}-t\right)}{2}\right] e^{-i \omega_{j+1} \left(t_{f}-t\right)} & \cos\left[ \frac{\Omega_{\mathrm{R}} \left(t_{f}-t\right)}{2} \right] e^{-i \omega_{j+1} \left(t_{f}-t\right)}
\end{array}
\right)
\end{equation}

\begin{figure}
\begin{center}
\includegraphics[width=0.75\textwidth]{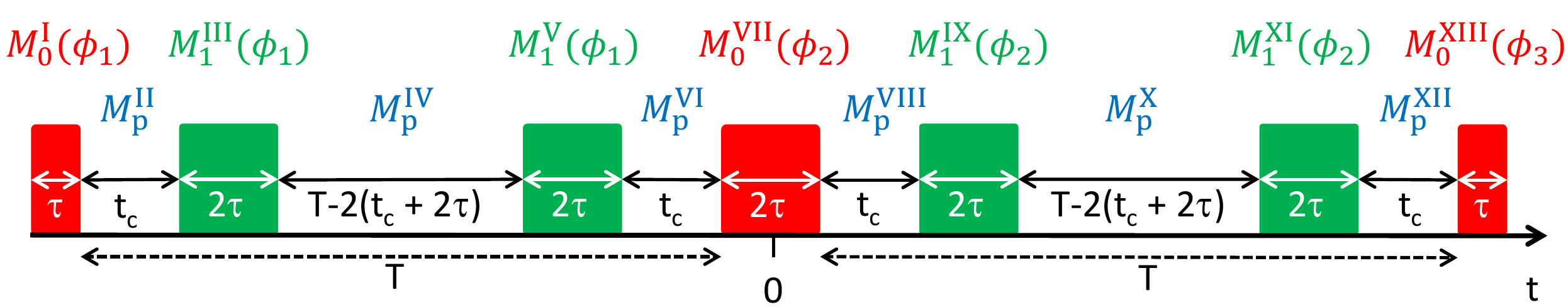}
\caption[figA1]{(Color online). Complete pulse sequence including the associated matrices for $N=1$. Red pulses correspond to the common diffracting pulses while green pulses correspond to accelerations produced on a single interferometric arm.\label{figA1}}
\end{center}
\end{figure}

\noindent where $\Omega_{\mathrm{R}}/\left(2 \pi\right)$ is the Rabi frequency, $\hbar \omega_{j}$ is the energy of the state $\left \vert j \right \rangle$, $\omega_{12}=\omega_{\mathrm{L1}}-\omega_{\mathrm{L2}}$ is the frequency difference between the two counter-propagating lasers, $t$ is the absolute timing at the start of the time step, $\phi$ is the laser phase at this instant and $t_{f}$ is the time at the end of the step. All other non-zero elements of $M_{j}$ lie on the diagonal and are simply $\left \lbrace e^{-i \omega_{q} \left(t_{f}-t\right)} \right \rbrace_{q \neq j,j+1}$. The interferometric sequence is then directly represented by a product of $5+8N$ of these matrices (see figure \ref{figA1}). These matrices depends on four parameters: the initial time $t$, the Rabi frequency $\Omega_{\mathrm{R}}$, the pulse duration $t_f-t$, and the laser phase difference $\phi$. As an example, the N=1 sequence corresponds to:

\begin{eqnarray}
\label{eq.annex.matprod}
\!\!\!\!\!\!\!\!\!\!\!\!\!\!\!\!\!\!\!\!\!\!\!\!\!\!\!\!\!\!\!\!\!\!\!\!M_{0}^{XIII}\left(T+\tau,\Omega_{\mathrm{R}},\tau,\phi_{3}\right).M_{p}^{XII}\left(T+\tau-t_{c},t_{c}\right).M_{1}^{XI}\left(T-\tau-t_{c},\Omega_{\mathrm{R}},2\tau,\phi_{2}\right). \nonumber\\
\!\!\!\!\!\!\!\!\!\!\!\!\!\!\!\!\!\!\!\!\!\!\!\!\!\!\!\!\!\!\!\!\!\!\!\!M_{p}^{X}\left(3\tau+t_{c},T-4\tau-2 t_{c}\right).M_{1}^{IX}\left(\tau+t_{c},\Omega_{\mathrm{R}},2\tau,\phi_{2}\right).M_{p}^{VIII}\!\!\left(\tau,t_{c}\right).M_{0}^{VII}\!\!\left(-\tau ,\Omega_{\mathrm{R}},2\tau,\phi_{2}\right).\nonumber\\
\!\!\!\!\!\!\!\!\!\!\!\!\!\!\!\!\!\!\!\!\!\!\!\!\!\!\!\!\!\!\!\!\!\!\!\!M_{p}^{VI}\left(-\tau-t_{c},t_{c}\right).M_{1}^{V}\left(-3\tau-t_{c},\Omega_{\mathrm{R}},2\tau,\phi_{1}\right).M_{p}^{IV}\left(-T+\tau+t_{c},T-4\tau-2 t_{c}\right).\nonumber\\
\!\!\!\!\!\!\!\!\!\!\!\!\!\!\!\!\!\!\!\!\!\!\!\!\!\!\!\!\!\!\!\!\!\!\!\!M_{1}^{III}\left(-T-\tau+t_{c},\Omega_{\mathrm{R}},2\tau,\phi_{1}\right).M_{p}^{II}\left(-T-\tau,t_{c}\right).M_{0}^{I}\left(-T-2\tau ,\Omega_{\mathrm{R}},\tau,\phi_{1}\right)
\end{eqnarray}

\noindent where the matrices are labelled with Roman numeral. For the light pulses matrices ($M_0$, $M_1$) the quadruplets $\left(t,\Omega_{\mathrm{R}},\left(t_{f}-t\right),\phi\right)$ are specified. During the free evolution ($\Omega_R =0$) the matrices ($M_{p}$) are determined by the doublets $\left(t,\left(t_{f}-t\right)\right)$. In this expression, we have used the usual convention of setting the phase of the 3-pulse interferometer to $\phi_{1,2,3}$. For clarity, we give also the full expression of the two matrices:

\begin{equation*}
\!\!\!\!\!\!\!\!\!\!\!\!\!\!\!\!\!\!\!\!\!\!\!\!\!\!\!\!\!\!\!\!\!\!\!\!M_{0}=
\footnotesize
\left(
\begin{array}{ccc}
\cos\left[\frac{\Omega_{\mathrm{R}} \left(t_{f}-t\right)}{2}\right] e^{-i \omega_{0} \left(t_{f}-t\right)} & -i e^{i \left(\omega_{12} t + n \phi\right)} \sin \left[\frac{\Omega_{\mathrm{R}} \left(t_{f}-t\right)}{2}\right] e^{-i \omega_{0} \left(t_{f}-t\right)} & 0 \\
- i e^{-i \left(\omega_{12} t + n \phi\right)} \sin\left[\frac{\Omega_{\mathrm{R}} \left(t_{f}-t\right)}{2}\right] e^{-i \omega_{1} \left(t_{f}-t\right)} & \cos\left[ \frac{\Omega_{\mathrm{R}} \left(t_{f}-t\right)}{2} \right] e^{-i \omega_{1} \left(t_{f}-t\right)} & 0 \\
0 & 0 & e^{-i \omega_{2} \left(t_{f}-t\right)}
\end{array}
\right)
\end{equation*}

and 

\begin{equation*}
\!\!\!\!\!\!\!\!\!\!\!\!\!\!\!\!\!\!\!\!\!\!\!\!\!\!\!\!\!\!\!\!\!\!\!\! M_{1}=
\footnotesize
\left(
\begin{array}{ccc}
e^{-i \omega_{0} \left(t_{f}-t\right)} &0 & 0 \\
0 & \cos\left[\frac{\Omega_{\mathrm{R}} \left(t_{f}-t\right)}{2}\right] e^{-i \omega_{1} \left(t_{f}-t\right)} & -i e^{i \left(\omega_{12} t + n \phi\right)} \sin \left[\frac{\Omega_{\mathrm{R}} \left(t_{f}-t\right)}{2}\right] e^{-i \omega_{1} \left(t_{f}-t\right)} \\
0 & - i e^{-i \left(\omega_{12} t + n \phi\right)} \sin\left[\frac{\Omega_{\mathrm{R}} \left(t_{f}-t\right)}{2}\right] e^{-i \omega_{2} \left(t_{f}-t\right)} & \cos\left[ \frac{\Omega_{\mathrm{R}} \left(t_{f}-t\right)}{2} \right] e^{-i \omega_{2} \left(t_{f}-t\right)} 
\end{array}
\right)
\end{equation*}

\noindent The output state corresponds to a superposition of $\left \vert 0\right \rangle$ and $\left \vert 1\right \rangle$, from which we find the accumulated phase $\Phi^{\left(N\right)}$, given by the usual formula:

\begin{equation}
\Phi^{\left(N\right)}=n \left(N+1\right)\left[\phi_{1}-2\phi_{2}+\phi_{3}\right]
\end{equation} 

\begin{figure}
\begin{center}
\includegraphics[width=0.40\textwidth]{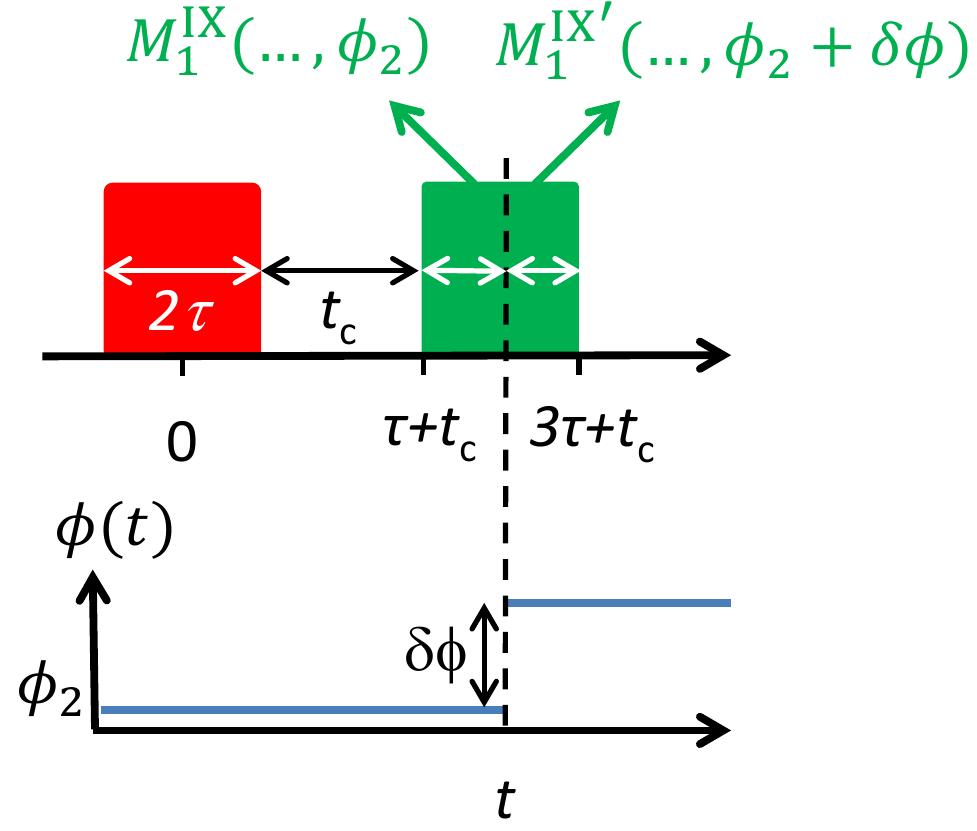}
\caption[figA2]{(Color online). Schematic representation of the phase jump $\delta \phi$ happening at time t during the IX-nth step. The associated matrix is split into two halves which depends on $t$.\label{figA2}}
\end{center}
\end{figure}

Finally, to obtain the sensitivity function (equation \ref{def-gphi}) one has to include a step change in phase for all possible times in the interferometer sequence. This can be done by splitting, depending on the time at which the step occurs, the corresponding matrix into two matrices with different laser phases (only relevant when $\Omega_{\mathrm{R}}\neq 0$). For example (see figure \ref{figA2})), if the phase step happens during the interval $\left[t_{c}+\tau;t_{c}+3\tau\right]$, one has to replace the term $M_{1}^{IX}\left(t_c+\tau ,\Omega_{\mathrm{R}},2\tau,\phi_{2}\right)$ by the product:

\begin{equation}
M_{1}^{IX'}\left(t,\Omega_{\mathrm{R}},3\tau+t_{c}-t,\phi_{2}+\delta \phi\right).M_{1}^{IX}\left(\tau+t_{c},\Omega_{\mathrm{R}},t-(\tau+t_{c}),\phi_{2}\right)
\end{equation}
\noindent and propagate the additional phase $\delta \phi$ in all subsequent laser phases present in later matrices. This results in an additional accumulated phase $\delta\Phi\left(\delta \phi,t\right)$ which directly gives the sensitivity function with:

\begin{equation}
g_{\phi}\left(t\right) = \lim \limits_{\delta \phi \rightarrow 0} \frac{\delta \Phi\left(\delta \phi,t\right)}{\delta \phi}
\end{equation}

\noindent where we have used the mid fringe assumption by choosing ($N=1$):

\begin{equation}
2 n \left(\phi_{1}-2\phi_{2}+\phi_{3}\right)=\frac{\pi}{2}
\end{equation}

The general form given by equation (\ref{gk}) was obtained from a full calculation on resonance ($\omega_{12} = \omega_{j+1}-\omega_{j}$ $\forall j$) for N=1,2 and 3 and then generalizing to arbitrary N using the temporal symmetry of the interferometer sequence.
\setcounter{section}{1} 
\label{app1}



\vspace{2pc}
\begin{thebibliography}{}
\bibitem{Borde89} Bord\'e C J 1989 \textit{Phys. Lett. A} \textbf{140} 10

\bibitem{FermiSchool} Tino G M and Kasevich M A, editors, Atom Interferometry 2014 \textit{Societ\`a Italiana di Fisica and IOS Press}, Amsterdam

\bibitem {louchet-chauvet_influence_2011} Louchet-Chauvet A, Farah T, Bodart Q, Clairon A, Landragin A, Merlet S and Dos Santos F P 2011 \textit{New J. Phys.} \textbf{6} 065025

\bibitem {hu_demonstration_2013} Hu Z K, Sun B L, Duan X C, Zhou M K, Chen L L, Zhan S, Zhang Q Z and Luo J 2013 \textit{Phys. Rev. A} \textbf{88} 043610

\bibitem {freier_mobile_2016} Freier C, Hauth M, Schkolnik V, Leykauf B, Schilling M, Wziontek H, Scherneck H G, M\"uller J, Peters A 2016 \textit{J. Phys.: Conf. Ser.} \textbf{723} 012050

\bibitem{Sorrentino_13} Sorrentino F, Bodart Q, Cacciapuoti L, Lien Y H, Prevedelli M, Rosi G, Salvi L, and Tino G M 2013 \textit{Phys. Rev. A} \textbf{89} 023607

\bibitem{Gustavson_00} Gustavson T L, Landragin A and Kasevich M A 2000 \textit{Class. QuantumGrav.} \textbf{17} 2385

\bibitem {Berg_composite_2015} Berg P, Abend S, Tackmann G, Schubert C, Giese E, Schleich W P, Narducci F A, Ertmer W and Rasel E , 2015 \textit{Phys. Rev. Lett.} \textbf{114} 063002

\bibitem {barrett_sagnac_2014} Barrett B, Geiger R, Dutta I, Meunier M, Canuel B, Gauguet A, Bouyer P and Landragin A 2014 \textit{C.R. Phys.} \textbf{15} 875

\bibitem {fixler_atom_2007} Fixler J B, Foster G T, McGuirk J M and Kasevich M A 2007  \textit{Science} \textbf{315} 5808: 74-77

\bibitem {rosi_precision_2014} Rosi G, Sorrentino F, Cacciapuoti L, Prevedelli M and Tino G M 2014 \textit{Nature} \textbf{510} 7506 518-521

\bibitem {bouchendira_new_2011} Bouchendira R, Clad\'e P, Guellati-Kh\'elifa S, Nez F, Biraben F 2011 Phys. Rev. Lett. 106, 08080

\bibitem{Parker_18} Parker R H, Yu C, Zhong W, Estey B and M\"{u}ller H 2018 \textit{Science} \textbf{360} 191-195 

\bibitem{Dimopoulos_08} Dimopoulos S, Graham P M, Hogan J M, Kasevich M A and Rajendran S 2008 \textit{Phys. Rev. D} \textbf{78} 122002

\bibitem {Zhou_test_2013} Zhou L \textit{et al} 2013 \textit{Phys. Rev. Lett.} \textbf{115} 013004 

\bibitem{Bonnin_13} Bonnin A, Zahzam N, Bidel Y and Bresson A 2013 \textit{Phys. Rev. A} \textbf{88} 043615

\bibitem{Barrett_16} Barrett B, Antoni-Micollier L, Chichet L, Battelier B, L\'{e}v\`{e}que T, Landragin A and Bouyer P 2016 \textit{Nat. Commun.} \textbf{7} 13786

\bibitem{Overstreet_18} Overstreet C, Asembaum P, Kovachy T, Notermans R, Hogan J M and Kasevich M A 2018  \textit{Phys. Rev. Lett.} \textbf{120} 183604

\bibitem{Wolf_07} Wolf P, Lemonde P, Lambrecht A, Bize S, Langevin A and Clairon A 2007 \textit{Phys. Rev. A} \textbf{75} 063608 

\bibitem{Jaffe_17} Jaffe M, Hasslinger P, Xu V, Hamilton P, Upadhye A, Elder B, Khoury J and M\"{u}ller H 2017 \textit{Nat. phys.} \textbf{13} 938-942 

\bibitem {canuel_matter-wave_2014} Canuel B, Amand L, Bertoldi A, Chaibi W, Geiger R, Gillot J, Landragin A, Merzougui M, Riou I, Schmid S P and Bouyer P 2014 \textit{E3S Web of Conference} \textbf{4} 01004

\bibitem {muller_atom_2008} M\"{u}ller H, Chiow S, Long Q, Herrmann S and  Chu S 2008 \textit{Phys. Rev. Lett.} \textbf{100} 180405

\bibitem {McDo_faster_2014} McDonald G D, Kuhn C C N, Bennetts S, Debs J E, Hardman K S, Close J D and Robins N P 2014 \textit{EPL} \textbf{105} 63001

\bibitem {clade_large_2009} Clad\'e P, Guellati-Kh\'{e}lifa S, Nez F and Biraben F 2009  \textit{Phys. Rev. Lett.} \textbf{102} 240402

\bibitem {chiow_102_2011} Chiow S, Kovachy T, Chien H C and Kasevich M A 2011 \textit{Phys. Rev. Lett.} \textbf{107} 130403

\bibitem{Borde_02} Bord\'e C J 2002  \textit{Metrologia} \textbf{39} 435-463 

\bibitem{Bongs_06} Bongs K, Launay R and Kasevich M A 2006 \textit{Appl. Phys. B} \textbf{84} 599-602 

\bibitem{Dubetsky_06} Dubetsky B and Kasevich M A 2006 \textit{Phys. Rev. A} \textbf{74} 023615

\bibitem{Kleinert_15} Kleinert S, Kajari E, Roura A and Schleich W P 2015 \textit{Phys. Rep.} \textbf{605} 1-50 

\bibitem{Dick_87}  Dick G J 1987 \textit{Proc. 19th Annu. Precise Time and Time Interval} \textbf{19} 133 

\bibitem{Cheinet08}  Cheinet P, Canuel B, Pereira Dos Santos F, Gauguet A, Yver-Leduc F and Landragin A 2008 \textit{IEEE Transactions on Instrumentations and Measurement} \textbf{57} 6, 1141

\bibitem{Muller08} M\"{u}ller H, Chiow S and Chu S 2008 \textit{Phys. Rev. A} \textbf{77} 023609 

\bibitem{Itano93} Itano W M, Bergquist J C, Bollinger J J, Gilligan J M, Heinzen D J, Moore F L, Raizen M G and Wineland D J 1993 \textit{Phys. Rev. A} \textbf{47} 3554 

\bibitem{Kasevich92} Kasevich M A and Chu S 1992 \textit{Appl. Phys. B} \textbf{54} 321 

\bibitem{Alan08} Alan J and Hui-Hui D 2008 \textit{Handbook of Mathematical Formulas and Integrals, 4\textsuperscript{th} edition}

\bibitem{Fang_18} Fang B, Mielec N, Savoie D, Altorio M, Landragin A and Geiger R 2018 \textit{New J. Phys.} \textbf{20} 023020 

\bibitem{Gauguet_08} Gauguet A, Mehlst\"{a}ubler T E, L\'ev\`eque T, Le Gou\"{e}t J, Chaibi W, Canuel B, Clairon A, Pereira Dos Santos F, Landragin A 2008 \textit{Phys. Rev. A} \textbf{78} 043615

\end{thebibliography}
\end{document}